\documentstyle[preprint,aps]{revtex}
\begin{document}

\draft

\title{Unified theory of strongly correlated electron systems}

\author{Yu-Liang Liu}
\address{Center for Advanced Study, Tsinghua University, Beijing 100084, 
People's Republic of China}

\maketitle

\begin{abstract}

In framework of eigen-functional bosonization method, we introduce an imaginary
phase field to uniquely represent electron correlation, and demonstrate that
the Landau Fermi liquid theory and the Tomonaga-Luttinger liquid theory can
be unified. It is very clear in this framework that 
the Tomonaga-Luttinger liquid behavior of one-dimensional
interacting electron gases originates from their Fermi structure, and the
non-Landau-Fermi liquid behavior of 2D interacting electron gases is induced
by the long-range electron interaction, while 3D interacting electron gases 
generally show the Landau Fermi liquid behavior.

\end{abstract}
\vspace{1cm}


\newpage
Since the discovery of high Tc cuprate superconductors\cite{1}, 
the strongly correlated electron systems has been extensively 
studied theoretically\cite{2,3,4,4a,4b,4c,5,6,7}. Now a common consensus
is reached that  
the low energy physics properties of 
cuprate superconductors, such
as anomalous normal state behavior and high superconducting transition
temperature, are determined by the strong electron correlation in their
copper-oxide plane(s).
However, up to now there is not a microscopic or 
phenomenal theory to successfully explain the normal and superconducting
physical properties of the cuprate superconductors, because one cannot
exactly and effectively treat the strong electron correlation of the systems.

It is well-known that usual metals can be described by the Landau Fermi
liquid theory\cite{8,9}, in which there is 
weak correlation among electrons, and
near the Fermi surface there exist well-defined quasi-particles (holes), thus 
the fundamental assumption of the Landau Fermi liquid theory is satisfied,
i.e., the states
of an interacting electron gas can be put into a one-to-one correspondence
via adiabatic continuation with those of the free electron gas. However, for
strong electron correlation, this fundamental assumption of the Landau Fermi
liquid theory may fail, and one does not have well-defined quasi-particles
(holes) near the Fermi surface. A one-dimensional interacting electron gas is
a good example, in which there is strong electron correlation even though for
small electron interaction, thus one does not have well-defined quasi-particles
(holes) near the Fermi levels $\pm k_{F}$ (its Fermi surface is composed of
two points, $\pm k_{F}$, defined by the Fermi momentum $k_{F}$).
It is described by the Tomonaga-Luttinger liquid theory\cite{10,11,12,13,14}, 
where in low
energy regime electron Green's function and other correlation functions 
present power-law behavior, and the correlation exponents are not universal,
and depend upon the electron interaction strength. Therefore, the
one-dimensional interacting electron gas is a strongly correlated electron
system even for weak electron interaction.

Generally, the low energy behavior of the one-dimensional (1D) interacting
electron gas is qualitatively different from that of three-dimensional (3D)
interacting electron gas, the former is represented by the
Tomonaga-Luttinger liquid theory, while the latter is represented by the
Landau Fermi liquid theory (see below). This difference derives from 
their different Fermi surface strctures. For 1D electron gas, its Fermi
surface is composed of two points $\pm k_{F}$, defined by the Fermi momentum
$k_{F}$, and its Hilbert space is drastically suppressed. There are only two 
kinds of elementary excitation modes, one is the excitation modes near these
two Fermi levels $\pm k_{F}$, respectively, where the energy spectrum is
approximately linear; and another one is the excitation modes with large
momentum ($2nk_{F}$, $n=1,2,...$) transfer between these Fermi 
levels $\pm k_{F}$, which usually make the system become an insulator ( for
a 1D electron gas this kind of excitation is absent). This drastically 
suppressed Hilbert space will induce the strong electron correlation even for
weak electron interaction, thus it is true that for a 1D interacting electron
gas the strong electron correlation originates from its special Fermi surface
structure, and the Tomonaga-Luttinger liquid theory is an universal theory of
1D electron gases. 
In contrast with the 1D electron gas, the Fermi surface of the 3D
electron gas is a sphere with a radius $k_{F}$, and its Hilbert space is 
enlarged comparing with that of the 1D electron gas. The low energy elementary
excitation modes are quasi-particles and quasi-holes. It is well-known that
even for long-range Coulomb interaction, the 3D electron gas
still can be described by the Landau Fermi liquid theory, where strong electron
interaction does not mean that there is the strong electron correlation, and 
it only produces weak and short-range electron correlation. Thus the Landau
Fermi liquid theory is an universal theory of 3D electron gases.

For 2D electron gas, the situation is different. Its Fermi surface is a circle
with a radius $k_{F}$, and for long-range Coulomb interaction,
it shows non-Landau-Fermi liquid behavior in the low energy regime\cite{15,16}.
We shall demonstrate that this anomalous behavior of the 2D interacting 
electron gas derives from the two aspects, one is its Fermi structure, and 
another one is the long-range electron interaction. These both effects induce
the strong electron correlation, thus the non-Landau-Fermi 
liquid behavior of the
2D strongly correlated electron gas cannot be completely represented by the
Tomonaga-Luttinger liquid theory. It shows not only some characters of the 
Landau Fermi liquid, but also some characters of the Tomonaga-Luttinger liquid.

It is very desirable to find an unified theory to represent not only 1D
interacting electron gases, but also 2D and 3D interacting electron gases. 
In the framework of the eigen-functional bosonization method\cite{17,18}, 
we try to give such the unified theory of strongly correlated electron gases,
which not only reduces to the Tomonaga-Luttinger liquid theory for the 1D 
interacting electron gas and to the Landau Fermi liquid thoery for the 3D 
interacting electron gas, but also can represent the low energy behavior of
the 2D interacting electron gas. We clearly demonstrate that
an imaginary phase field which naturally appears 
in the eigen-functional bosonization, is a key parameter field
hiden in the strongly correlated electron systems and 
represents the electron
correlation, and the eigen-functionals can be used to define the states of
the interacting electrons. By calculation the overlap of these 
eigen-functionals with the eigen-functions of the free electrons, we can
judge whether or not there exist well-defined quasi-particles (holes) near the 
Fermi surface. The imaginary phase field is a very important quantity in our
representation, and it
determines the low energy behavior of the system. Moreover, as $d\geq 2$
it does not contribute to the action of the system.

In general, we consider the Hamiltonian
(omitting spin label $\sigma$ of the electron operators $c_{k}$)
\begin{equation}
H=\sum_{k}\epsilon(k)c^{\dagger}_{k}c_{k}+\frac{1}{2L^{d}}\sum_{q}
v(q)\rho(q)\rho(-q)
\label{1}\end{equation}
where $\epsilon(k)=k^{2}/(2m)-\mu$, $\mu$ is the chemical potential,
$d$ is the dimension number, $\rho(q)=(1/L^{d})\sum_{k}c^{\dagger}_{k}
c_{k+q}$ is the electron density operator, and $v(q)$ is the Coulomb potential.
The last term represents usual electron's Coulomb interaction (four-fermion
interaction). Usually, one can 
introduce the Hubbard-Stratonovich (HS) field $\phi(x,t)$ to descouple this 
four-fermion interaction term. With the standard path integral method\cite{19},
we have the action of the system,
\begin{eqnarray}
S &=& \displaystyle{ \int dt\int d^{d}x\left\{\Psi^{\dagger}(x,t)\left[
i\frac{\partial}{\partial t}+\mu+\frac{\nabla^{2}}{2m}-\phi(x,t)\right]
\Psi(x,t)\right\}} \nonumber \\
&+& \displaystyle{\frac{1}{2TL^{d}}\sum_{q,\Omega}\frac{1}{v(q)}
\phi(-q,-\Omega)\phi(q,\Omega)}
\label{2}\end{eqnarray}
where $\Psi(x,t)$ is the electron field.
The system now reduces into that the electrons move in the 
HS field $\phi(x,t)$, and we have the eigen-functional equation,
\begin{equation}
\left[i\frac{\partial}{\partial t}+\mu+\frac{\nabla^{2}}{2m}-\phi(x,t)
\right]\Psi_{k\omega}(x,t,[\phi])=E_{k\omega}[\phi]\Psi_{k\omega}(x,t,[\phi])
\label{3}\end{equation}
where the eigen-value $E_{k\omega}[\phi]=\omega-\epsilon(k)-\Sigma_{k}[\phi]$,
and the self-energy $\Sigma_{k}[\phi]=\int^{1}_{0}d\xi\int dt d^{d}x \phi(x,t)
\Psi^{\dagger}_{k\omega}(x,t,[\xi\phi])\Psi_{k\omega}(x,t,[\xi\phi])$ 
is a regular function, and independent of $\omega$.
The eigen-functionals $\Psi_{k\omega}(x,t,[\phi])$ can be used to define the
elementary low energy excitation modes, i.e., quasi-particles (holes) and/or
collective excitation modes, and have the formal solutions,
\begin{equation}
\Psi_{k\omega}(x,t,[\phi])=A_{k}\left(\frac{1}{TL^{d}}\right)^{1/2}
e^{Q_{k}(x,t)}e^{i{\bf k}\cdot{\bf x}-i(\omega-\Sigma_{k}[\phi])t}
\label{4}\end{equation}
where $|A_{k}|\sim 1$ is the normalization constant, and $T\rightarrow\infty$ 
is the time length of the system. These eigen-functionals are composed of two 
parts, one presents the free electrons, and another one presents the 
correlation of the electrons produced by the electron interaction.
Thus we can formally write
\begin{eqnarray}
\Psi_{k\omega}(x,t,[\phi])=\psi_{k\omega}(x,t)e^{Q_{k}(x,t)}
\nonumber\end{eqnarray}
The phase fields $Q_{k}(x,t)$ determine
the low energy behavior of the system, and satisfy the usual Eikonal
equation with the condition $Q_{k}(x,t)=0$ as $\phi(x,t)=0$,
\begin{equation}
\left[i\frac{\partial}{\partial t}+i\frac{{\bf k}\cdot{\bf \nabla}}{m}+
\frac{\nabla^{2}}{2m}\right]Q_{k}(x,t)-\phi(x,t)=-\frac{1}{2m}
\left({\bf \nabla}Q_{k}(x,t)\right)^{2}
\label{5}\end{equation}
which can be exactly solved by a series expansion of the HS field $\phi(x,t)$
and/or by computer calculations\cite{20,21}.

With the solution of the eigen-functionals $\Psi_{k\omega}(x,t,[\phi])$
(\ref{4}), the action of the system reads (omitting constant terms),
\begin{eqnarray}
S &=& \displaystyle{
\frac{1}{2TL^{d}}\sum_{q,\Omega}\frac{\phi(-q,-\Omega)\phi(q,\Omega)}
{v(q)}} \nonumber \\
&-& \displaystyle{\frac{1}{2}\int dt d^{d}x \phi(x,t)\left[ F_{1}(x,t,\delta)
+F_{2}(x,t)\right]_{{\bf \delta}\rightarrow 0}}
\label{6}\end{eqnarray}
where $F_{1}(x,t,\delta)=-(1/(2\pi)^{d})\int d^{d}k\theta(-\epsilon(k))
\sin({\bf k}\cdot{\bf \delta}){\bf \delta}\cdot{\bf \nabla}Q^{I}_{k}(x,t)$,
$F_{2}(x,t)=(2/(2\pi^{d}))\int d^{d}k\theta(-\epsilon(k))Q^{R}_{k}(x,t)$,
and $\theta(x)$ is a step function, i.e., $\theta(x)=1$ for $x>0$ and 
$\theta(x)=0$ for $x<0$.
We have written the phase field as $Q_{k}(x,t)=Q^{R}_{k}(x,t)+iQ^{I}_{k}(x,t)$,
a real part $Q^{R}_{k}(x,t)$ and an imaginary part $Q^{I}_{k}(x,t)$. We shall
demonstrate that the phase field $Q^{I}_{k}(x,t)$ completely determine the low
energy behavior of the system. It is also noted that the 
action of the system is composed of two parts, one is from the real phase
field $Q^{R}_{k}(x,t)$, and another one is from the imaginary phase field
$Q^{I}_{k}(x,t)$. Due to the Fermi surface structure of the system, the 
momentum integral in $F_{1}(x,t,\delta)$ can be written as\cite{22},
$\int d^{d}k=S_{d-1}\int d|k| |k|^{d-1}\int^{\pi}_{0}d\theta 
(\sin\theta)^{d-2}$, where $S_{d}=2\pi^{d/2}/\Gamma(d/2)$. 
As $d\geq 2$, the integration of $\sin({\bf k}\cdot{\bf \delta})$ is regular, 
thus as ${\bf 
\delta}\rightarrow 0$ the function $F_{1}(x,t,\delta)=0$, the imaginary
phase filed $Q^{I}_{k}(x,t)$ has no contribution to the action of the system.
Only at $d=1$, it has contribution to the action, where the real phase
field $Q^{R}_{k}(x,t)$ is zero (see below). This property is independent of 
the electron interaction, it is completely determined by the Fermi surface
structure of the system.

We now consider the 1D interacting electron gas, in which the Fermi surface
is composed of two Fermi levels, $\pm k_{F}$, and the electron energy
spectrum can be written as, $\epsilon(k)=\pm v_{F}k$, where $v_{F}$ is the 
electron Fermi velocity. The branch $\epsilon(k)=v_{F}k$ represents the 
right-moving electrons, and the branch $\epsilon(k)=-v_{F}k$ represents the
left-moving electrons. In general, the electron interaction term
reads, $(1/L)\sum_{q}V(q)\rho_{R}(q)\rho_{L}(-q)$, where $\rho_{R(L)}(q)$
are the right- and left-moving electron densities, respectively, and
$V(q)\sim V$, a constant. To decouple this four-fermion interaction, we 
introduce two HS fields $\phi_{R}(x,t)$ and $\phi_{L}(x,t)$, and have
two sets of eigen-functionals representing the right- and left-moving
electrons, respectively,
\begin{eqnarray}
\Psi_{Rk\omega}(x,t,[\phi]) &=& \displaystyle{ \left(\frac{1}{TL}\right)^{1/2}
e^{Q_{R}(x,t)}e^{ikx-i(\omega-\Sigma_{R}[\phi])t}} \nonumber \\
\Psi_{Lk\omega}(x,t,[\phi]) &=& \displaystyle{ \left(\frac{1}{TL}\right)^{1/2}
e^{Q_{L}(x,t)}e^{ikx-i(\omega-\Sigma_{L}[\phi])t}}
\label{7}\end{eqnarray}
where $\Sigma_{R(L)}[\phi]$ is a regular quantity, and independent of $k$
and $\omega$. The phase fields $Q_{R}(x,t)$ and $Q_{L}(x,t)$ are 
independent of $k$, and satisfy the simplified Eikonal equation\cite{17,23},
\begin{eqnarray}
\displaystyle{
\left(i\frac{\partial}{\partial t}+iv_{F}\frac{\partial}{\partial x}\right)
Q_{R}(x,t)-\phi_{R}(x,t)} &=& 0 \nonumber \\
\displaystyle{
\left(i\frac{\partial}{\partial t}-iv_{F}\frac{\partial}{\partial x}\right)
Q_{L}(x,t)-\phi_{L}(x,t)} &=& 0
\label{8}\end{eqnarray}
These linear differential equations can be easily solved, and the phase
fields $Q_{R}(x,t)$ and $Q_{L}(x,t)$ are imaginary because the HS fields
$\phi_{R(L)}(x,t)$ are real. It is worthy noted that the imaginary phase 
fields $Q_{R(L)}(x,t)$ not only determine the electron correlation, but also
contribute to the action of the system. This is qualitatively different from
that in 2D and 3D electron gases, where the imaginary part of the phase
field $Q_{k}(x,t)$ does not contribute to the action due to their Fermi
surface structures.

It is very simple to prove that the 1D interacting electron gas is a strongly
correlated system even for very weak electron interaction $V\sim 0$.
With the action (\ref{6}), by the simple calculation we can obtain the 
relations,
\begin{eqnarray}
<\psi_{R(L)k\omega}(x,t)\Psi^{\dagger}_{R(L)k\omega}(x',t,[\phi])>_{\phi}
&\sim & \displaystyle{\left(\frac{1}{L}\right)^{\alpha}e^{ik(x-x')}} 
\label{9} \\
<e^{Q_{R(L)}(x,t)}e^{-Q_{R(L)}(x',t)}>_{\phi} &\sim &
\displaystyle{ \left(\frac{1}{|x-x'|}\right)^{2\alpha}, \;\;\; |x-x'|
\rightarrow\infty}
\nonumber\end{eqnarray}
where $\alpha\sim (1-V/(2\pi\hbar v_{F}))/2$ for $V\sim 0$,
is the dimensionless coupling strength parameter,
the eigen-functions $\psi_{R(L)k\omega}(x,t)$ present the right(left)-moving 
free electrons,
and $<...>_{\phi}$ means taking functional average over the HS field 
$\phi_{R(L)}(x,t)$.
The first equation presents the zero overlap between the eigen-functionals
$\Psi_{R(L)k\omega}(x,t,[\phi])$ of the interaction electrons and the 
eigen-functions of the free electrons ($V=0$) as $L\rightarrow\infty$, thus
the states of the interacting electron gas does not have one-to-one
correspondence via adiabatic continuation with those of the free electron
gas, and there are not well-defined quasi-particles (holes) near its two
Fermi levels $\pm k_{F}$. The second equation presents the strong electron
correlation, in which the electron correlation length is infinity even for
weak electron interaction, this can be easily seen by re-writing
the right side of the last
equation in (\ref{9}) as, $\exp\{-2\alpha\ln|x-x'|\}$.
Due to this strong electron correlation, the low energy excitation modes of the
1D interacting electron gas are those collective excitation modes, such as
the charge and spin density waves, and there are not well-defined 
quasi-particles (holes) near the two Fermi levels $\pm k_{F}$. 
Thus the Tomonaga-Luttinger
liquid theory is a strongly correlated theory, and 
is universal for 1D interacting electron gases.

As $d\geq 2$, the real phase field $Q^{R}_{k}(x,t)$ is finite, and the 
imaginary phase field $Q^{I}_{k}(x,t)$ does not contribute to the action of
the system. For simplicity, we can solve the Eikonal equation by
neglecting the quadratic term $({\bf \nabla}Q_{k})^{2}$. This approximation
is reasonable because for the long-range Coulomb interaction, only the states
near the Fermi surface with momentum $q<q_{c}$ ($q_{c}\ll k_{F}$)
are important in the low energy regime, and for the smooth function
$Q_{k}(x,t)$ this quadratic term is proportional to $(q_{c}/k_{F})^{2}\sim 0$.
With this approximation, we can obtain the simple effective action,
\begin{equation}
S_{eff.}=\frac{1}{2TL^{d}}\sum_{q,\Omega}\left(\frac{1}{v(q)}-\chi(q,\Omega)
\right)|\phi(q,\Omega)|^{2}
\label{11}\end{equation}
where $\chi(q,\Omega)$ is usual Lindhard function\cite{19}. In fact, the above
approximation is equivalent to usual random-phase approximation (RPA). 
However, in present framework, it gives more useful and important informations
than usual RPA perturbation method, because we have not only the real
phase field, but also the imaginary phase field. This imaginary phase field
does not contribute to the action of the system, but it determines the 
electron correlation. For a 3D electron gas with long-range Coulomb interaction
$v(q)=e^{2}/(4\pi q^{2})$, by simple calculation, we have the relations,
\begin{eqnarray}
\displaystyle{
<\psi_{k\omega}(x,t)\Psi^{\dagger}_{k\omega}(x',t,[\phi])>_{\phi}} &\sim & 
\displaystyle{Z_{k}e^{i{\bf k}\cdot({\bf x}-{\bf x'})}} 
\nonumber \\
\displaystyle{
<e^{iQ^{I}_{k}(x,t)}e^{-iQ^{I}_{k}(x',t)}>_{\phi}} &\sim & 
\displaystyle{ e^{z^{I}_{k}(x-x')}}
\label{12}\end{eqnarray}
where $Z_{k}$ is finite, and $z^{I}_{k}(x)$ is a smooth function. As $|{\bf x}|
\rightarrow\infty$, the function $z^{I}_{k}(x)$ goes to zero. The first
equation presents that the eigen-functionals $\Psi_{k\omega}(x,t,[\phi])$
have strong overlap with the eigen-functions $\psi_{k\omega}(x,t)$, and the 
second equation presents that there is weak electron correlation even for
long-range Coulomb interaction. Thus the fundamental assumption of the
Landau Fermi liquid theory is satisfied, 
thus there are well-defined quasi-articles
(holes) near the Fermi surface. The Landau Fermi liquid theory is universal
for 3D interacting electron gases. Therefore, we can generally conclude that 
there does not exist non-Landau-Fermi liquid behavior in the 3D electron gases
as $q_{c}\ll k_{F}$.
For a 2D electron gas with the long-range Coulomb interaction $v(q)=e^{2}/
(4\pi q^{2})$, the situation is different from that in the 3D electron gas.
By simple calculation, we can obtain the relations,
\begin{eqnarray}
\displaystyle{ <\psi_{k\omega}(x,t)\Psi^{\dagger}_{k\omega}(x',t,[\phi])
>_{\phi}} &\sim & \displaystyle{ \left(\frac{1}{q_{c}L}\right)^{\beta}
e^{i{\bf k}\cdot({\bf x}-{\bf x'})}} \label{13} \\
\displaystyle{ <e^{iQ^{I}_{k}(x,t)}e^{-iQ^{I}_{k}(x',t)}>_{\phi}} &\sim &
\displaystyle{e^{-2\beta\ln(q_{c}|{\bf x}-{\bf x'}|)}, \;\;\;
|{\bf x}-{\bf x'}|\rightarrow\infty}
\nonumber\end{eqnarray}
where $\beta=e^{2}/(2(4\pi)^{2}\omega_{p})$ is the dimensionless coupling 
strength parameter, where $\omega_{p}$ is the plasma frequency.
In fact, $q_{c}L\rightarrow\infty$, the first equation presents that the 
eigen-functionals $\Psi_{k\omega}(x,t,[\phi])$ 
has no (or infinitesimal) overlap with the eigen-functions 
$\psi_{k\omega}(x,t)$,
and the second equation presents the strong electron correlation. In this case,
the fundamental assumption of the Landau Fermi liquid theory fails, and
the system shows the non-Landau-Fermi liquid behavior. Even though
the asymptotic behavior of the single electron Green's function is similar
to that of the 1D interacting electron gas (see, (\ref{9}) and (\ref{13})),
while they originate from different physics. The former is producted by the
plasmon excitation of the system, while the latter is induced by the gapless
charge (spin) density wave(s). Thus this
non-Landau-Fermi liquid behavior is different from the usual Tomonaga-Luttinger
liquid behavior, because the former is induced by the strong long-range
electron interaction, while the latter is induced by the Fermi structure of
the 1D electron gases. However, this difference can also be clearly seen from 
their effective actions of the density field. For 1D interacting electron 
gases, this action can be easily obtained by usual bosonization 
method\cite{13,14}, 
where the density field has well-defined propagator (density
wave). For the 2D electron gas with long-range Coulomb interaction, the 
effective action (\ref{11}) can be written as with the density field,
\begin{equation}
S_{eff.}[\rho]=\frac{1}{2TL^{2}}\sum_{q,\Omega}\left(\frac{1}{\chi(q,
\Omega)}-v(q)\right)|\rho(q,\Omega)|^{2}
\label{14}\end{equation}
where the density field does not have well-defined propagator. 
It was shown in Ref.\cite{16} that as $e^{2}\gg 4\pi$ the system can be 
described as a Landau Fermi liquid formed by chargeless quasi-particles 
which has vanishing wavefunction overlap with the bare electrons in the system.
In the short range $\xi\sim 1/q_{c}$, one can also define local quasi-particles
(holes). For short range and/or weak electron interaction, the 2D
interacting electron gases would show the Landau Fermi liquid behavior in the
low energy region.

In summary, in the simple framework of the eigen-functional bosonization 
method, we have demonstrated that the imaginary phase field $Q^{I}_{k}(x,t)$
can completely represent the electron correlation, and
the Landau Fermi liquid theory and the
Tomonaga-Luttinger liquid theory can be unified.
It is very clear in this framework that: 
a). the Tomonaga-Luttinger liquid behavior
of 1D interacting electron gases originates from their Fermi structure which
is composed of two Fermi levels $\pm k_{F}$; the phase field is imaginary, and
contributes to the action of the systems. 
b). the non-Landau-Fermi liquid behavior
of 2D interacting electron gas is induced by the long-range Coulomb 
interaction. The imaginary phase field $Q^{I}_{k}(x,t)$ does not contribute
to the action of the system. However, for other anomalous gauge 
interaction, the system can also show the non-Landau-Fermi liquid behavior
if the imaginary phase field $Q^{I}_{k}(x,t)$ satisfies the relations in 
(\ref{13}). Generally, 2D interacting electron gases may show the Landau
Fermi liquid behavior or non-Landau-Fermi liquid behavior in the low energy
region, which completely depends upon the electron interaction. 
c). 3D interacting
electron gases generally show the Landau Fermi liquid behavior. The electron 
interaction cannot alter this property as $q_{c}\ll k_{F}$, and it only
modifies the weak electron correlation length and the overlap 
weight with the bare electron wavefunctions. 

This method can also be used to treat a 2D electron gas with transverse
gauge interaction, and gives the same results as that in Ref.\cite{27,28}.
However, this method can clearly show that the non-Fermi liquid behavior
of this system originates from the imaginary phase field which is a function
of transverse gauge fields.

We thank T. K. Ng for helpful discussions.

\end{document}